\documentstyle[12pt,aaspp4]{article}
\begin{document}
\title{Temporal Variability of the X-ray Emission of the Crab Nebula Torus}
\author{C. Greiveldinger \& B. Aschenbach}
\affil{Max-Planck-Institut f\"ur Extraterrestrische Physik }
\authoraddr{Giessenbachstrasse 1, D-85740, Garching, Germany}

\begin{abstract}
We present ROSAT HRI observations of the Crab Nebula spanning the six
years from 1991 March to 1997 March.  A comparison of the observations
reveals that there are significant ($> 10 \sigma$) differences in the
emission from the nebula at a rate of $\sim 2\%/$year.  These
differences are confined to rather large ($\gtrsim 25\arcsec \times 25
\arcsec $), well-defined regions.  One region is coincident with the
end of the southern polar outflow, while the others are coincident
with the torus surrounding the pulsar.  The various regions have
different time histories, but they generally show a monotonic change
in count rate, some increasing and others decreasing.  A possible
explanation for the observed behavior is that the radial bulk motion
of the relativistic wind after shock passage has slowed by 10 to 20 \%
over the six years of observations.
\end{abstract}
\keywords{ISM: individual (Crab Nebula)---supernova remnants---X-rays: ISM}

\section{Introduction}

For years, the Crab Nebula has provided astronomers with numerous
puzzles, and it continues to do so to this day.  The Crab presents an
excellent opportunity to study the interaction of a pulsar with its
surrounding supernova remnant, and X-rays have provided some of the
most probing insights.

Early in the era of X-ray astronomy, it was learned that the X-ray
emission from the Crab was not confined to a point source and that the
centroid of the X-ray emission was not coincident with the pulsar
(e.g. Bowyer et al.\markcite{bea64} 1964).  To explain this offset,
Aschenbach \& Brinkmann\markcite{ab75} (1975) proposed a model for the
distribution of X-rays around the pulsar.  This model built on
previous models (e.g. Rees \& Gunn\markcite{rg74} 1974) that were
based on the idea that the pulsar at the center of the nebula released
a fraction of its energy in the form of a relativistic wind of
electrons and positrons.  The seemingly strange emission geometry
could be reproduced if this wind was preferentially directed in the
pulsar's rotational equatorial plane.  When the ram pressure of the
wind equals the pressure of the surrounding medium, a shock occurs,
and the particles begin to synchrotron radiate.  This scenario gives
rise to a torus of emission in the rotational equatorial plane of the
pulsar.  Indeed, with imaging X-ray telescopes, this proposed
morphology was detected (Brinkmann, Aschenbach, \&
Langmeier\markcite{bal85} 1985).

This much alone can explain extended emission, but it does not
necessarily explain why the brightest region of X-ray emission was to
the northwest of the pulsar.  Aschenbach \& Brinkmann (1975) further
suggested that this brighter region could be due to a local
enhancement of the magnetic field.  The pulsar's proper motion is also
towards the northwest (Trimble\markcite{t71} 1971), and it is
suggested that the pulsar is moving relative to the nebula.  If this
is the case, it will ``pile-up'' magnetic field lines in the forward
direction, producing a region of stronger magnetic field and higher
synchrotron emission.

An alternative explanation for the enhanced northwest brightness is
provided by relativistic beaming effects (Pelling et
al.\markcite{pea87} 1987).  Using scanning modulation collimator
observations from 22 to 64 keV, Pelling et al. (1987) found that the
X-ray emission to the northwest was 5 times brighter than to the
southeast.  They were able to explain this difference with
relativistic effects if the bulk motion of the radiating particles is
at a speed $v/c = \beta \approx 0.3$, which is in good agreement with
expectations for the conditions achieved after a highly relativistic
wind passes through a shock (Blandford \& Rees\markcite{br74} 1974).

In this letter, we present the most detailed and deepest X-ray data
set available for the Crab Nebula to date and discuss some striking
variations that occur over a span of 6 years.  In Section 2, we
discuss the observations and processing techniques.  In Section 3, we
present the regions where the strongest variations occur and show
light curves, and in Section 4, we discuss the plausibility of certain
models in light of the new data.  Finally, Sections 5 and 6 present
some remaining difficulties, summarize the results, and make basic
predictions.

\section{Observations and Processing}

All observations discussed herein were taken with the R\"ontgen
Satellite (ROSAT) High Resolution Imager (HRI)
(Tr\"umper\markcite{t83} 1983, Pfeffermann et al.\markcite{p86} 1986),
and processing was performed with the MIDAS/EXSAS software package
(Zimmermann et al.\markcite{zea93} 1993).  Table~\ref{tab:obs-data}
gives details of observation dates and exposure times.  All
observations were performed with the satellite in wobble mode, so any
particular feature in the nebula passed over several hundred detector
pixels.  In this way, the effects of pixel to pixel variations in the
detector are minimized.  In addition, the observation from September
1992 was actually a series of calibration observations with the pulsar
offset from the detector center along various directions.  These
observations were aligned and co-added to produce a single image for
1992 September.  Further processing was performed on the 1995 March
and the 1996 March observations to improve the image quality.  These
two observations suffered from inaccuracies in the satellite aspect
solution that resulted in the pulsar image appearing quite noticeably
elliptical, and this effect was corrected.

Each data set was originally binned at $1 \arcsec$ resolution (twice
the detector pixel size, but still less than the mirror resolution).
These images were aligned by performing a two dimensional Gaussian fit
to the pulsar image, and the centers determined by these fits were
aligned.  The formal errors of the center position from the Gaussian
fits were significantly less than $1 \arcsec$.  After this, a
cross-correlation analysis was performed on the images to determine if
there was a rotation about the pulsar position that would produce a
better alignment.  Since it has the longest exposure time, we used the
1997 image as the reference.  We found that only the 1991 and 1992
images showed an improved correlation with a rotation of $2\arcdeg$.
This relative rotation is presumably due to a spacecraft rotation, and
is not an intrinsic property of the nebula.  In the remainder of this
letter, when referring to the 1991 and 1992 images, we are discussing
the images after this rotation correction has been applied.  This
rotation does not affect our results in any significant manner.

After the images were aligned, they were then rebinned in $5 \arcsec$
pixels to better match the telescope resolution
(Aschenbach\markcite{a88} 1988).  The rebinning was done so that the
Gaussian center of the pulsar emission was itself centered on a pixel.
This resulted in most of the pulsar emission falling in a single
pixel.  All further analysis was performed at this $5 \arcsec$
resolution.

\section{Comparison of Observations}
Since we were dealing with observations of varying exposure times, we
had to be careful in how to compare the different images.  Using the
count rate would be a logical solution, but although the first four
observations (1991--1996) had count rates that were within 2\% of each
other ($\sim 235$ c/s, not corrected for dead time), the 1997
observation had a count rate that was $\sim 11\%$ lower.  The cause of
this discrepancy is not clear, although we attribute it to a change in
the HRI sensitivity and not to an intrinsic variation in the nebula.
We decided to normalize the images by dividing by the number of
counts within the region that contained the nebula.  Since the
nebular emission is not symmetric with respect to the pulsar, this
region is not quite centered on the pulsar.  We chose a $280\arcsec
\times 280\arcsec$ (56 $\times$ 56 pixel) region that ranges in right
ascension from $\alpha = 05^{\rm h}34^{\rm m}21\fs0$ to $05^{\rm
h}34^{\rm m}41\fs2$ and in declination from $\delta = +21\arcdeg
58\arcmin 28\farcs5$ to $+22\arcdeg 3 \arcmin 8\farcs5$.

We chose $\chi^2$ statistics to quantify the significance of detected
variations between images $A$ and $B$, and with the normalization that
we have adopted
\begin{equation}
\chi_i^2 = \frac{(N_{i,A}/N_{{\rm tot},A} - N_{i,B}/N_{{\rm tot},B})^2}
{N_{i,A}/N_{{\rm tot},A}^2(1 + N_{i,A}/N_{{\rm tot},A})
+ N_{i,B}/N_{{\rm tot},B}^2(1 + N_{i,B}/N_{{\rm tot},B})},
\label{eqn:chi2}
\end{equation}
where $N_i$ is the number of photons in the $i{\rm th}$ pixel, and
$N_{\rm tot}$ is the number of counts in the region of interest.  The
terms in parentheses in the denominator are negligible for single
pixel comparisons, but later we will select regions larger than one
pixel, and this additional term becomes more important.

With five different observations, there are ten different image
comparisons that can be computed.  For each of the possible
combinations, we produced a $\chi^2$ image, where the content of a
given pixel is determined by the formula give above.  These images are
displayed in Figure~\ref{fig:allchis}.  Each row and column
corresponds to a given observation date, and the image at the
intersection of a row and column is the $\chi^2$ image derived from
the two corresponding observations.  The reduced $\chi^2$ value of the
entire region is also indicated at the lower left of each image. In
this figure, the scale shown at the bottom, ranges from $\chi_i^2 =
41$ to 100, where 41 is a $5\sigma$ difference when taking into
account the number of image pixels (3136).

Two things are immediately noticeable.  First, as one compares images
separated by longer intervals, the differences become more strongly
pronounced.  This is also presented in Figure~\ref{fig:chi2_vs_t}.
Second, the differences are confined to very particular regions in the
image; they are not smoothly distributed over the nebula.
Figure~\ref{fig:chi9792} shows the comparison of the 1992 and 1997
images (the two with the longest exposure times).  Overlaid on the
$\chi^2$ image are contours indicating the position of the pulsar and
the outer regions of the X-ray structure, indications of the positions
of the torus and jets, and outlines of the most pronounced regions of
change.

By choosing regions larger than one pixel, we can dramatically improve
the statistics.  This has been done for the regions indicated in
Figure~\ref{fig:chi9792}.  The light curves and $\chi^2$ values for
these regions are presented in Figure~\ref{fig:lightcurves}.  Another
interesting result is that these regions do not show the same history.
Some regions show an increase in emission, while others show a
decrease.  Even the two most significant regions (W and SW in
Figure~\ref{fig:chi9792}), which are adjacent to one another, show
opposite behavior.

Region SW appears to be a feature projecting radially from the pulsar.
If this is truly the case, one would expect that regions at different
radial distances would respond at different times to a change
originating in the vicinity of the pulsar.  The light curves of five
regions along the length of SW (Figure~\ref{fig:swlc}) reveal that
this is not the case.  The onset of the brightness increase appears to
set in around 1994 in all the light curves.  This can be understood if
regions SW is actually composed of several areas in the torus with
equal radial distances from the pulsar.  Any changes originating near
the pulsar would reach these locations at the same time, and the
apparent radial orientation would simply be a projection effect.

\subsection{Systematic Errors}

With the large count rate of the Crab Nebula, the statistical errors
turn out to be quite small, so we have to be very careful of possible
systematic errors.  There are two major possible sources of error:
variations in detector sensitivity over the six years and misalignment
of the images.  

To test for detector variations, we performed the identical analysis
on two archival observations of Cas A.  The first was taken on 29 July
1990 and had an exposure of 8 ksec.  The second was taken from 23
December 1995 to 1 February 1996 and lasted 180 ksec.  The images were
aligned and rebinned in a manner similar to that described above.
Since Cas A lacks a bright point source, our alignment is likely to be
less accurate than that for the Crab.  Even with these larger
positioning uncertainties, we calculate that over the entire region
that we have considered, $\chi_{\nu}^2 \approx 2$ for Cas A, which is
much less than the values calculated for the Crab Nebula at similar
time separations (Figure~\ref{fig:chi2_vs_t}).  The same thing is true
in each of the regions identified in Figure~\ref{fig:chi9792}.  Thus,
using Cas A as our standard we conclude that the observed effects are
not caused by variations in detector sensitivity.

To test for possible errors in our alignment procedure we used the
1997 Crab image and shifted it by 1\arcsec, 2 \arcsec, and 5\arcsec \ in
various directions, binned it to 5\arcsec \ resolution, and compared
it with the original 1997 image.  We found that an offset of even 1\arcsec
\ could produce an extremely high $\chi_{\nu}^2$.  The largest
$\chi_{\nu}^2 \approx 4$.  Inspection of the resulting $\chi^2$ images
reveals that the region near the pulsar produces the largest
contribution to the total $\chi^2$, a behavior that is not seen in our
data.  If we exclude the $3 \times 3$ pixel region
centered on the pulsar, the $\chi_{\nu}^2$ drops by $\geq 1.5$ in our
deliberately shifted images, while excluding the same region in our
data produces changes in $\chi_{\nu}^2 \leq 1$.

In addition, other factors lead us to believe that poor alignment does
not contribute strongly.  To reproduce the temporal behavior seen in
Figure~\ref{fig:chi2_vs_t} and Figure~\ref{fig:lightcurves}, the
relative offsets could not simply be random.  The $\chi^2$ in the
shifted images is largest near the pulsar, as mentioned above, or at
surface brightness edges.  Although we do see some variations occuring
near the edges of the nebula, the most striking regions are within or
close to the torus, away from areas of strong brightness contrast.
Therefore, we also believe that our alignments are accurate enough
that the significance of these variations is not in doubt.

\section{Discussion}

A most intriguing feature of these variations is that, with one
exception (region E) they occur in regions that are either aligned
with the torus or with a polar jet.  Here we will discuss the regions
that are apparently associated with the torus (S, N, W, NE, SW).

As mentioned above, there have been two ideas put forth to explain the
enhanced nebular brightness to the northwest: increased magnetic field
in the region and relativistic effects.  We will first try to
understand the observed variations in terms of these models under the
assumption that the intrinsic electron and photon spectra have
remained constant.

\subsection{Magnetic Field Pile-up}

Let us first consider the magnetic field explanation.  As pointed out
in Aschenbach \& Brinkmann (1975), Shklovskii\markcite{s57} (1957)
shows that increasing the magnetic field by a factor $m$ increases the
synchrotron volume emissivity by $m^{\Gamma+1}$, where $\Gamma$ is
the power law index of the electron distribution in the nebula.  For
synchrotron emission, the observed index of the differential photon
energy spectrum ($\alpha$) can be related to the index of the
differential electron spectrum ($\Gamma$) by $\Gamma = 2\alpha - 1$
(e.g. Rybicki \& Lightman\markcite{rl79} 1979).  Using $\alpha = 2$
for the Crab implies that $\Gamma = 3$, and the corresponding increase
in synchrotron volume emissivity is given by $m^4$.  If we let this
magnetic field ``boost'' factor be represented by $b_B = m^4$, then we
can determine that $\frac{\Delta b_B}{b_B} = 4\frac{\Delta m}{m}$.
Assuming that the length of a given emitting region along the line of
sight remains unchanged implies that the emitting volume is constant.
Therefore changes in the volume emissivity will be directly observable
as changes in the surface brightness.

Although it is not straightforward to specify a change in observed
brightness for every region (in Figure~\ref{fig:lightcurves}, regions
N and S are particularly confusing), we can assign a value of
$\frac{\Delta b_B}{b_B} = -0.13$ for region W, which means that
$\frac{\Delta m}{m} = -0.033$.  Following the reasoning of Aschenbach
\& Brinkmann (1975), an increase in the magnetic field strength could
be due to the piling up of magnetic field lines as the pulsar moves
through the nebula.  The direction of the pulsar's motion (Trimble
1971) is roughly towards region W, so with this explanation, one would
expect this region to show an increase in count rate.  In fact, it
shows the opposite.  To produce a decrease in emission, the magnetic
field lines would have to spread farther apart.  Furthermore, to
produce adjacent regions that exhibit opposite brightness changes
would require a rather complicated magnetic field reorientation.

\subsection{Relativistic Effects}

The suggestion by Pelling et al. (1987) that the brightness
enhancement to the northwest is due to a combination of relativistic
aberration and Doppler shifting can also be investigated.  They
determined that, by using the torus orientation determined by
Aschenbach \& Brinkmann (1975), the increased brightness to the
northwest can be explained if the emitting particles have a bulk
motion with $v = 0.3c$.  

We can in principle perform the same analysis with the ROSAT HRI data.
It is not obvious, though, how to select regions on the torus for
comparison.  If the bulk flow is uniform and is confined to a plane,
our $\beta$ determination should not depend on which regions are
chosen.  Unfortunately, we would still have the difficulty that
different sections of the torus show different time behaviors in
addition to the fact that the better angular resolution reveals that
certain regions are really a combination of polar and toroidal
emission.  

The torus appears to be inclined at 30\arcdeg\ to the line of sight
with the major axis of the projected toroidal ellipse aligned $\approx
45\arcdeg$ east of north (Aschenbach \& Brinkmann 1975, Hester et
al.\markcite{hea95} 1995).  The northwest region is pointed toward the
observer, while the southeast region points away.  In order to be able
to determine an estimate of $\beta$ for the bulk motion we chose six
regions aligned with the torus (three on the front side and three on
the back).  To limit the effects of the variations, we used the sum of
all five images to determine the brightness.  All regions were
20\arcsec $\times$ 20\arcsec, and Table~\ref{tab:beta} indicates the
locations of the regions ($\Phi$ is the angle measured
counterclockwise from west) and the number of photons in the region.

We will represent the apparent boost that power-law photons receive
from relativistic effects by $b_r$, where
\begin{equation}
b_r = \frac{(1 - \beta^2)^{3/2}}{(1 - \beta \cos \theta)^3}.
\label{eqn:rel_boost}
\end{equation}
Here $\theta$ is the angle (measured in the observer's frame) between
the direction of the electron bulk motion and the direction of
radiation, and we have assumed a value $\alpha = 2$ for the photon
spectral index.  For the geometry described above, $\cos \theta = \cos
30 \cos (\Phi - 45)$.  Table~\ref{tab:beta} also indicates the values
for $\theta$ at the various torus positions.

To calculate $\beta$, one needs to consider the relative brightnesses
of the regions.  There are difficulties when one tries to compare
region $b$ to region $c$ and region $e$ to region $f$.  According to
this simple model, $b$ and $c$ (and also $e$ and $f$) should be
essentially identical regions, since they are roughly equally spaced
from the toroidal axis on the front (rear) side of the torus.
Inspection of Table~\ref{tab:beta} shows that they are not the same,
and trying to solve for $\beta$ for these combinations produces
negative results.  Therefore, we disregard these values and average
the rest to determine $\bar \beta = 0.25$, which we adopt for the
remainder of this discussion.  Assuming that the injected electron
spectrum remains constant, there are two ways to produce changes in
the relativistic boost $b_r$, namely change $\theta$ or change
$\beta$.  We will investigate each of these possibilities.


A change in $\theta$ really would mean that the orientation of the
torus had changed or that the orientation of the injected relativistic
wind had changed.  For simplicity, we will only consider the case
where the torus remains rigid and centered on the pulsar.  The
projection of the torus onto the plane of the sky is characterized by
two angles.  One angle determines the azimuthal orientation of the
major axis of the projected ellipse.  Varying this angle corresponds
to a rotation of the projected ellipse in the plane of the sky.
During our image alignment we would have removed any rotations of this
sort, and therefore a variation of the azimuthal orientation of the
projected ellipse cannot be a cause of the observed changes.

The angle that is critical for changing the boosting factor is the
angle of inclination between the torus and our line of sight.  We will
refer to the angle as $\psi$ and note that we have used $\psi = 30
\arcdeg$ in previous discussions.  If we allow $\psi$ to vary, this
will produce a change in $\theta$ determined by $\cos \theta = \cos
\psi \cos (\Phi - 45)$.  Substituting this value for $\cos \theta$
into Equation~\ref{eqn:rel_boost} and differentiating, we find that
the fractional change in the relativistic boost is given by
\begin{equation}
\frac{\Delta b_r}{b_r} = \frac{-3 \beta \sin \psi \cos (\Phi - 45)}
	{1 - \beta \cos \psi \cos (\Phi -45)}\Delta \psi.
\label{eqn:delta_theta}
\end{equation}
Assuming that changes in $\psi$ and $b_r$ are small, we can use $\psi
= 30 \arcdeg$ and $\beta = 0.25$ to calculate the necessary change in
$\psi$ to produce the observed variation in brightness in a given
region.  If we try this for region W (from
Table~\ref{tab:beta_change}, $\Phi = 10 \arcdeg$, $\Delta b/b \approx
-0.13$) we find that $\Delta \psi = 14 \arcdeg$.  This value may at
first seem quite large, but one must recall that the torus has a width
of 15\arcdeg \ to 20\arcdeg \ (Aschenbach \& Brinkmann 1975, Hester et
al. 1995).  Thus, it is possible to imagine that the plane of the
wind's bulk motion is varying at an amplitude that is comparable to the
width of the torus.

If we extend this analysis to other regions, we encounter
difficulties---most notably at region SW and region NE.  Since these
two regions essentially lie along the major axis of the projected
ellipse we should see very little variation in their brightness if
$\psi$ changes.  In fact, for SW, we actually see the greatest amount
of change, and the $\Delta \psi$ required to produce it is
unreasonably large.  For this reason, it seems unlikely that the
changes observed are due to variations in the orientations of the wind
plane.

Now, let us consider the effects of a changing $\beta$ on $b_r$.
If we follow the same procedure as used above, we find that
\begin{equation}
\frac{\Delta b_r}{b_r} = \frac{3\beta}{1-\beta^2} 
	\left[\frac{\cos \theta - \beta}{1 - \beta \cos \theta}\right]
	\frac{\Delta \beta}{\beta} = Q(\theta,\beta)
	\frac{\Delta \beta}{\beta}.
\label{eqn:delta_beta}
\end{equation}

From Figure~\ref{fig:lightcurves}, we estimate the fractional change
in the brightness ($\Delta b/b$) of each of the torus regions
(Table~\ref{tab:beta_change}).  For regions N and S, these values
should only be taken as very rough estimates, since their behavior is
not well-defined.  Using these values and the values for $\theta$
indicated in Table~\ref{tab:beta} we calculate the required fractional
change in $\beta$, and these values are presented in
Table~\ref{tab:beta_change}.  We should point out that there is no
solution for $\Delta \beta/\beta$ if $\cos \theta = 0.25$
(i.e. $\theta \approx 76 \arcdeg$, meaning $\Phi = 118 \arcdeg$ or
$332 \arcdeg$).  At these locations $\Delta b_r = 0$, regardless of
the change in $\beta$, and furthermore, for a given change in $\beta$
these angles define regions where $\Delta b_r$ changes sign (see
Figure~\ref{fig:explanation}).  Another point to note is that the
calculated value of $\Delta \beta/\beta$ is very sensitive to the
angle used when $\theta \approx 90 \arcdeg$.

All the regions except region S require a decrease in the value of
$\beta$.  Region S may be complicated by the fact that it lies near
the southern outflow, and so it is likely a combination of toroidal
and jet emission.  The same thing could be said about region N and
region W, but in the north, the rotational pole is likely to point
away while the torus is inclined toward the observer.  Beaming effects
should mean that the torus dominates in the north, and, likewise, the
jet could be dominant in the south.  Region SW is also problematic,
because $\Delta \beta/\beta \approx -1$, using $\theta = 90 \arcdeg$.
Using other values of $\theta$ near 90\arcdeg \ only makes matters
worse.

Keeping these difficulties in mind, it is still intriguing that the
computed $\Delta \beta/\beta$ is the same order of magnitude for
regions at various locations on the torus that have different
magnitudes and different directions of brightness change.  At least in
a qualitative manner, a decrease in $\beta$ of $\sim$20\% can explain
the changes seen here.

\subsection{Spectral Changes}

The possibilities discussed above have neglected possible changes in
the photon or electron spectra.  Such changes could be caused by 
variations in the emitted flux, changes in the spectral index,
differences in absorbing neutral hydrogen along the line of sight, or
a combination of all three.  If one keeps the unabsorbed energy flux
in the ROSAT HRI energy band constant, assumes $N_H = 3 \times
10^{21}$ cm$^{-2}$, and simply varies the photon index, the photon
flux can vary sharply.  Changing the photon index $\alpha = 2.0$ by
$\pm 0.2$ results in $\sim 15\%$ changes in the observed HRI count
rate.  If we fix $\alpha$ and vary $N_H$ we find that to produce a
10\% change in count rate, the $N_H$ must change by $\sim 1 \times
10^{21}$ cm$^{-2}$.  So, it would be possible to vary the spectral
parameters to produce a change in the observed count rate, but, once
again, the orchestration of these parameter changes as a function of
position seems to require some complicated maneuvering to match the
observations.

\section{Problems}

Although it seems to us that the best way to explain our observations
is that $\beta$ of the bulk particle motion has changed by $\sim 20\%$
from 1991 to 1997, there still remain some difficulties with this
explanation.  First, if $\beta$ has changed by so much, we might
expect to be able to repeat our $\beta$ determination for the 1991 and
1997 observations and detect a difference.  Unfortunately, for reasons
discussed above, determining $\beta$ is not a straightforward process.
If we do apply the method previously described, we compute $\beta
\approx 0.26 \pm 0.13$ for both 1991 and 1997, where the quoted error
is the standard deviation.  Any 20\% variation is lost in the large
uncertainty.


It is also not clear how such a large change in $\beta$ would be
produced.  An intriguing possibility would be that this is related to
a decrease in the speed of the relativistic wind produced by the
pulsar.  Several authors have estimated that the maximum speed that
the pulsar can impart to the electrons varies with the pulsar period
(e.g. \markcite{gj69}Goldreich \& Julian 1969, \markcite{og69}Ostriker
\& Gunn 1969).  These are pre-shock speeds, and our observations are
sensitive to the electrons after they have passed through the shock.
Therefore, we have to relate the post-shock $\beta$ to the pre-shock
$\beta$.  For ultra-relativistic motions, Blandford \& Rees (1974)
have shown that
\begin{equation}
\frac{v_{\rm post}}{v_{\rm pre}} = \frac{1 + 3p_{\rm pre}/p_{\rm post}}
	{3 + p_{\rm pre}/p_{\rm post}},
\label{eqn:shock_condition}
\end{equation}
where $v$ is the velocity and $p$ is the pressure in the indicated
region.  For a strong shock $v_{\rm post}/v_{\rm pre} = \onethird$.
If this is the only effect on the pulsar wind, then, with $\beta_{\rm
post} = 0.25$, that would indicate that $\beta_{\rm pre} = 0.75$,
which is significantly less than expected.  Moreover, a fractional
change in $\beta_{\rm post}$ requires the same fractional change in
$\beta_{\rm pre}$.  If this were true, then $\beta_{\rm pre}$ would
have to have dropped to $\approx 0.6$, meaning that the energy
contained in the particle wind had decreased by $\sim 17\%$.  Even if
only a small fraction of the pulsar's spin-down luminosity is fed into
the relativistic wind, this is still a considerable change.

Instead of originating from the pulsar, the observed changes could
stem from variations in the shock boundary.  This would eliminate the
need for such a large change in the pre-shock relativistic wind speed.
In addition, it could provide a simple explanation for the observed
spread in the derived $\Delta \beta/\beta$ for the various regions, a
quantity which is related to the magnitude of brightness variation, as
well as the time scale similarities between the regions.  The
magnitude of the variations could be dominated by local parameters
such as density, pressure, or magnetic field strength, while the time
scale for variation is a global property that depends on the speed at
which a perturbation is transmitted.  This speed could conceivably be
the relativistic wind, sound, or Alfv\'{e}n speed.

\section{Summary and Predictions}

With ROSAT HRI observations from 1991 to 1997, we have detected
significant but localized changes in the X-ray emission of the Crab
Nebula.  If the increased brightness of the northwest portion of the
torus is due to relativistic beaming, the variations can be explained,
in a general sense, as being due to a $\sim 20\%$ decrease in the
post-shock bulk motion.  Why there should be such a large decrease is
not known at this time, but it seems unlikely that it can be explained
by a decrease in the $\beta$ of the pre-shock relativistic wind.

We expect that this decrease in $\beta$ cannot continue for long.  If
it maintains this rate of decrease, $\beta$ will drop to zero rather
rapidly.  It seems likely that what we are observing is some transient
behavior of the nebula, and that there is a ``recovery'' mechanism
which will bring the $\beta$ factor back to a larger value.  With
continuing observations we will be able to better understand the
nature of these changes and the properties of the pulsar wind and its
shock.

\section{Acknowledgments}
CG would like to thank J. Tr\"umper and W. Becker for helpful
discussions and MPE for its excellent hospitality.

\newpage
\begin{figure}
\caption{The $\chi^2$ images produced from the observations.  Each row
and column corresponds to a particular observation, and the
intersection of a row and column is the comparison of those two
observations.  The number in the lower left corner of each frame is
the $\chi_{\nu}^2$ for that image, and the ROSAT HRI X-ray image of
the Crab Nebula in the lower left is presented to orient the reader.
Each image is $280\arcsec \times 280\arcsec$.  The intensity scale at
the bottom ranges from $\chi^2 = 41 (5\sigma)$ to 100.}
\label{fig:allchis}
\end{figure}

\begin{figure}
\caption{Plot of reduced $\chi_{\nu}^2$ as a function of time
separation.  The crosses represent the Crab Nebula data shown in
Figure~\ref{fig:allchis}.  The square is the result of comparing two
Cas A observations used to estimate systematic errors.  As
can be seen, the Cas A comparison does not show changes as large as
those detected for the Crab Nebula at similar time separations.}
\label{fig:chi2_vs_t}
\end{figure}

\begin{figure}
\caption{The 1997 vs. 1992 $\chi^2$ image.  These observations have
have two of the longest exposures, and their comparison shows the most
differences.  Indicated are the regions we chose for further study,
outlines of the pulsar, torus, and jet locations, and one of the
outermost nebular contours.  These outlines are also indicated in
Figure~\ref{fig:explanation} and are only meant to indicate the
orientation of the image.  The image intensity is the same as that in
Figure~\ref{fig:allchis}.}
\label{fig:chi9792}
\end{figure}

\begin{figure}
\caption{Lightcurves and $\chi^2$ values for the regions indicated in
Figure~\ref{fig:chi9792}.  The ordinate for the lightcurves is the
percent of emission that a particular region contributes to the total
nebular emission, and the errorbars are smaller than the symbol sizes.}
\label{fig:lightcurves}
\end{figure}

\begin{figure}
\caption{Light curves from locations along the length of region SW.
Each light curve was determined for a $10\arcsec \times 10\arcsec$
region at the indicated angular distance from the pulsar.  The
apparently simultaneous onset of brightness increase near 1994 can be
understood if these regions are all at the same radial distance from
the pulsar and the apparent radial alignment is merely a projection
effect.}
\label{fig:swlc}
\end{figure}

\begin{figure}
\caption{Image of the Crab Nebula presented with 1\arcsec \ binning.
Panel A is displayed with a linear intensity scale, and overlaid are
the locations of the pulsar, torus, and jets.  Panel B is the same
image displayed with a logarithmic scaling.  The additional radial
lines in this panel indicate how the relativistic boost ($b_r$) should
change for a decrease in $\beta$ (see text for details).  Also
displayed are the observed brightness changes for the torus regions.}
\label{fig:explanation}
\end{figure}

\vfill
\eject

\begin{deluxetable}{cccc}
\tablecaption{ROSAT HRI Observation Dates and Exposure Times.
\label{tab:obs-data}}
\tablecolumns{4}
\tablehead{
\colhead{Year}&\colhead{Start Date}&\colhead{End Date}&
\colhead{Exposure (ks)}}
\startdata
1991 & 20 Mar & 24 Mar & 12 \nl
1992 & 16 Sep & 17 Sep & 27 \nl
1995 & 4 Mar & 15 Mar & 8 \nl
1996 & 9 Mar & 29 Mar & 33 \nl
1997 & 7 Mar & 19 Mar & 40 \nl
\enddata
\end{deluxetable}

\begin{deluxetable}{cccc}
\tablecaption{Locations of regions used to determine $\beta$ of
post-shock wind. The top three regions are on the front side of 
the torus, while the bottom three are on the back side. \label{tab:beta}}
\tablecolumns{4}
\tablehead{
\colhead{Region}&\colhead{$\Phi$ (degrees)}&\colhead{$\theta$ (degrees)}&
\colhead{$N_{\rm photons}$} }

\startdata
$a$ & 55  &  30 & $1.6725 \times 10^6$ \nl
$b$ & 105 & 65  & $1.4917 \times 10^6$ \nl
$c$ & 355 &  55 & $1.1398 \times 10^6$ \nl

\tablevspace{0.5cm}

$d$ & 235 & 150 & $5.003 \times 10^5$ \nl
$e$ & 275 & 125 & $7.218 \times 10^5$ \nl
$f$ & 175 & 125 & $5.489 \times 10^5$ \nl
\enddata
\end{deluxetable}

\begin{deluxetable}{cccccc}
\tablecaption{Regions of observed brightness
variations. \label{tab:beta_change}}
\tablecolumns{6}
\tablehead{
\colhead{Region}&\colhead{$\Phi$ (degrees)}&\colhead{$\theta$
(degrees)}& \colhead{$\left(\frac{\Delta b}{b}\right)$\tablenotemark{a}}
&\colhead{$\frac{1}{Q(\theta,\beta)}$\tablenotemark{b}}&
\colhead{$\frac{\Delta \beta}{\beta}$} }

\startdata
W  &  10 &  46 & -0.13 & +2.3 & -0.3  \nl
N  &  80 &  46 & -0.04 & +2.3 & -0.1  \nl
NE & 135 &  90 & +0.08 & -5.0 & -0.4  \nl
S  & 280 & 120 & -0.04 & -1.8 & {\bf +0.1} \nl
SW & 315 &  90 & +0.20 & -5.0 & -1.0 \nl
\enddata
\tablenotetext{a}{Estimated from Figure~\ref{fig:lightcurves}.}
\tablenotetext{b}{From Equation~\ref{eqn:delta_beta}.}
\end{deluxetable}

\end{document}